\documentclass[amsmath,amssymb,reprint,bibnotes]{revtex4-2}
\usepackage{graphicx}
\usepackage{bm}
\usepackage{amsmath} 
\usepackage{graphics}
\usepackage[dvipsnames]{xcolor}
\usepackage[mathlines]{lineno}
\usepackage{braket}
\usepackage{subfigure}
\usepackage[compact]{titlesec}
\begin{document}
\title{Dynamic control of dipole decay rate via graphene plexcitons}
\author{Hira Asif$^{\bf (1,2)}$}
\author{Taner Tarik Aytas$^{\bf (1,3)}$}
\author{Ramazan Sahin$^{\bf (1,2)}$}\email{rsahin@itu.edu.tr}

\affiliation{${\bf (1)}$ {Department of Physics, Akdeniz University, 07058 Antalya, Turkey}}
\affiliation{${\bf (2)}$ {Türkiye National Observatories, TUG, 07058 Antalya, Turkey}}
\affiliation{${\bf (3)}$ {Türkiye National Observatories, DAG, 07058 Erzurum, Turkey}}

\date{\today}

\begin{abstract}
Active control of the radiative properties of quantum emitters through engineered light–matter interactions is a key challenge in nanophotonics and quantum optics. In this work, we demonstrate dynamic modulation of dipole's decay rate by exploiting the tunable plexcitonic modes (graphene plasmons and QD-excitons) in the strong coupling regime. By integrating a quantum dot inside a graphene spherical shell and tuning the local optical response of hybrid modes via voltage-bias, we achieve continuous and reversible control over the decay rate, leading to significant enhancement or suppression of dipole emission from near- to far-infrared regime. Furthermore, the plexcitonic peaks shows much sharper linewidths in contrast to bare graphene plasmons even in the off-resonant coupling which indicates higher sensitivity of the systems at tuned wavelengths. We demonstrate the phenomenon with the numerical solution of 3D Maxwell's equations using MNPBEM tool. Our approach demonstrate a versatile platform for programmable emission control and offer a promising pathway for developing reconfigurable quantum photonic devices, such as tunable single-photon sources and ultrafast optical switches.
\end{abstract}

\maketitle
 
\section{Introduction}
Controlling the radiative properties of quantum emitters (QE) through engineered nanostructures lies at the heart of modern nanophotonics and quantum optics \cite{Cai2018,Schrinner2020}. Such control offers wide range of applications in emerging quantum technologies, including on-chip light source, quantum information processing and nanoscale sensing \cite{Alfieri2023,Clarke2024}. In particular, the spontaneous emission rate of a quantum emitter can be significantly modified by engineering the local density of states (LDOS) in its environment. This offers a route to enhancing emission in a controlled manner, thereby enabling deterministic light-matter coupling at the nanoscale \cite{Muravitskaya2022}.\\
Plasmonic systems have long been explored for emission rate control due to their ability to confine electromagnetic fields well below the diffraction limit. However, conventional metal-based plasmonic platforms suffer from high ohmic losses and lack active tunability, limiting their applicability in dynamic and programmable photonic devices. Recent advances in two dimensional (2D) plasmonic materials, particularly graphene, have opened new avenues in this direction \cite{Cuevas2016}. Graphene plasmons offer exceptional mode confinement, broadband tunability via electrostatic gating, and longer propagation lengths in the mid-infrared to terahertz regimes \cite{Fei2012, Chen2012, Fang2013, Gullans2013}. In recent years, electric tuning of graphene has demonstrated its potential use in vast applications such as sensing \cite{Li2014}, metasurfaces \cite{Balci2018}, switching \cite{Ju2011} and strong light-matter coupling \cite{Koppens2011}.\\
In addition, exciton resonances in quantum dots offers a uniquely strong light-matter interaction that is spectral tunable which opens new pathways to explore active/tunable optical elements and nanophotonic devices. However, external tuning mechanisms such as electric or magnetic fields and strain provide only limited shifts in exciton energy, keeping their operation within a narrow spectral range \cite{Lu2017,Stier2016}.
\begin{figure}
\includegraphics[scale=0.24]{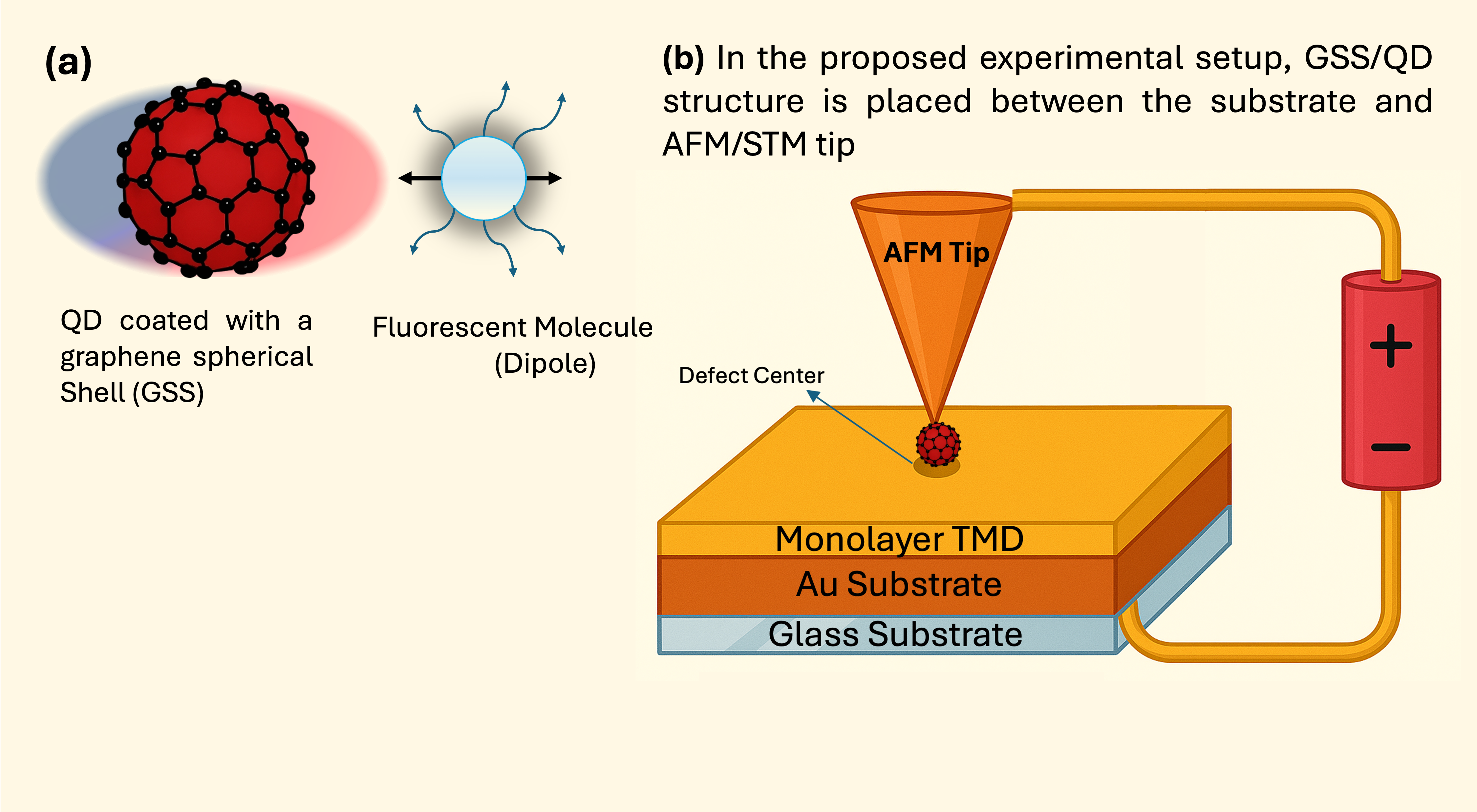}
\caption{\label{fig:1} (a) Interaction of graphene plexcitons supported by graphene plasmons of GSS and excitons of QD with a florescent molecule. (b) Illustration of ultrafast photonic sensor operating in infrared regime.}
\end{figure} 
While exciton-polariton extend the tunability, they remain constrained by the Rabi splitting. An even more promising approach arises from the formation of plexcitons , i.e., hybrid quasiparticles resulting from the strong coupling between plasmons and excitons. These modes combine the advantages of both constituents the field localization of plasmons and tunability of excitons with the coherence and narrow linewidths. The optical properties of these plexcitons significantly modifies the spectral dynamics of molecular system placed in the vicinity of these structures. Furthermore, to achieve coherent control between molecule and plexcitonic modes, the choice of materials and their combinations is crucial importance. Materials must be selected not only for their individual characteristics but also for their chemical and structural compatibility when stacked. For instance, combining graphene with TMDs leverages graphene’s high mobility and TMDs’ strong light-matter interactions, leading to hybrid structures with enhanced functionality. While strong coupling between plasmon-QD systems has been extensively studied in the context of Rabi splitting and energy transfer, its potential for actively modulating the decay rates of dipole molecule in the off-resonant regime remains largely unexplored.\\ 
In this work, we investigate the active modulation of decay rate of a dipole emitter through voltage tunable plexcitonic modes formed by coupling graphene plasmons with quantum dot (QD). We show that  the hybrid plexcitonic modes strongly coupled showing resonances with sharp linewidths even in the off-resonant regime. The spectral dynamics of these plexitonic states can be continuously modulated, over a broad spectral of infrared regime, by changing the chemical potential of graphene through external voltage bias. When quantum emitter (QE) in the form of dipole is placed in the vicinity of plexcitons, its temporal dynamics shifts significantly yielding enhancement or suppression of spontaneous emission. Moreover, the sharp linewidth of graphene plexcitons reveal high spectral sensitivity and dynamic control in quantum photonic applications.\\ 
 \begin{figure}[t!]
      \centering
        \includegraphics[scale=0.44]{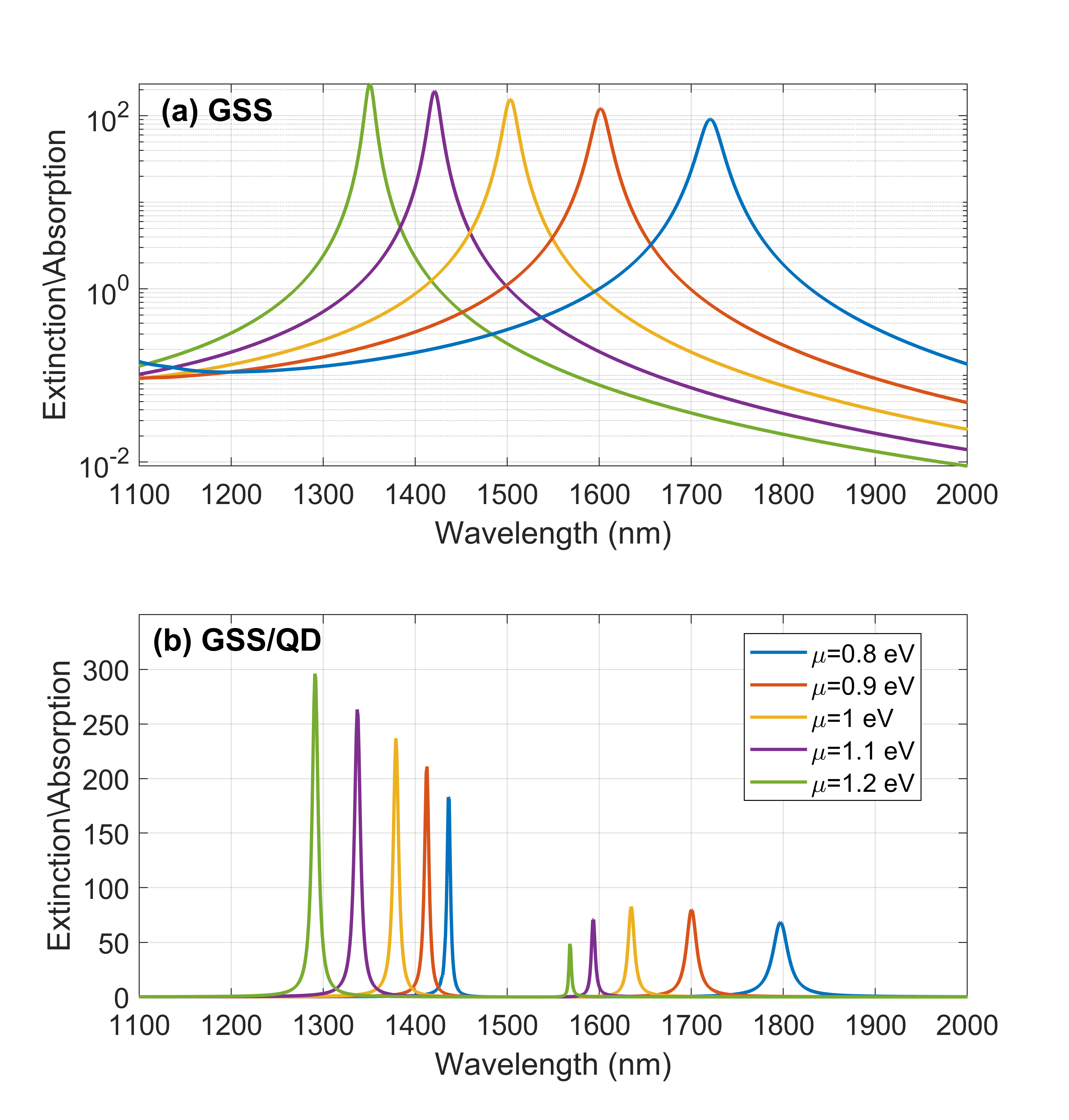}
    \caption{\label{fig:2} (a,b) Extinction/absorption spectrum of (a) GSS and (b) GSS/QD for different chemical potential of graphene as a function of excitation wavelength.}
\end{figure}
\subsection{Electromagnetic Simulations of QD Coated with Graphene}
Our model system compose of a QD (exciton) coated with a graphene spherical shell (GSS) with diameter 10 nm placed between the substrate and an AFM tip, as shown in Fig.\ref{fig:1}. Such type of structure has been demonstrated experimentally \cite{Christensen2015}. The AFM tip is attached to a voltage bias through layered structure of Au and monolayer TMDs on SiO$_2$. The layer of TMDs is source of dipole emitter which decay rate dynamics is probed by graphene plexcitons. To evaluate the extinction/absorption spectra, we perform 3D electromagnetic simulations, based on the MNPBEM toolbox to analyze the voltage-dependent spectral features of plexciton and corresponding decay rate control of emitter \cite{Hohenester2012}. In boundary element simulations, the dielectric permittivity of graphene spherical shell is defined as \cite{Vakil2011}, 
\begin{equation}
\epsilon(\omega) = 1+\frac{4\pi\sigma(\omega)}{\omega d}
 \label{eq:1}
\end{equation}
\begin{figure}[t!]
      \centering
        \includegraphics[scale=0.27]{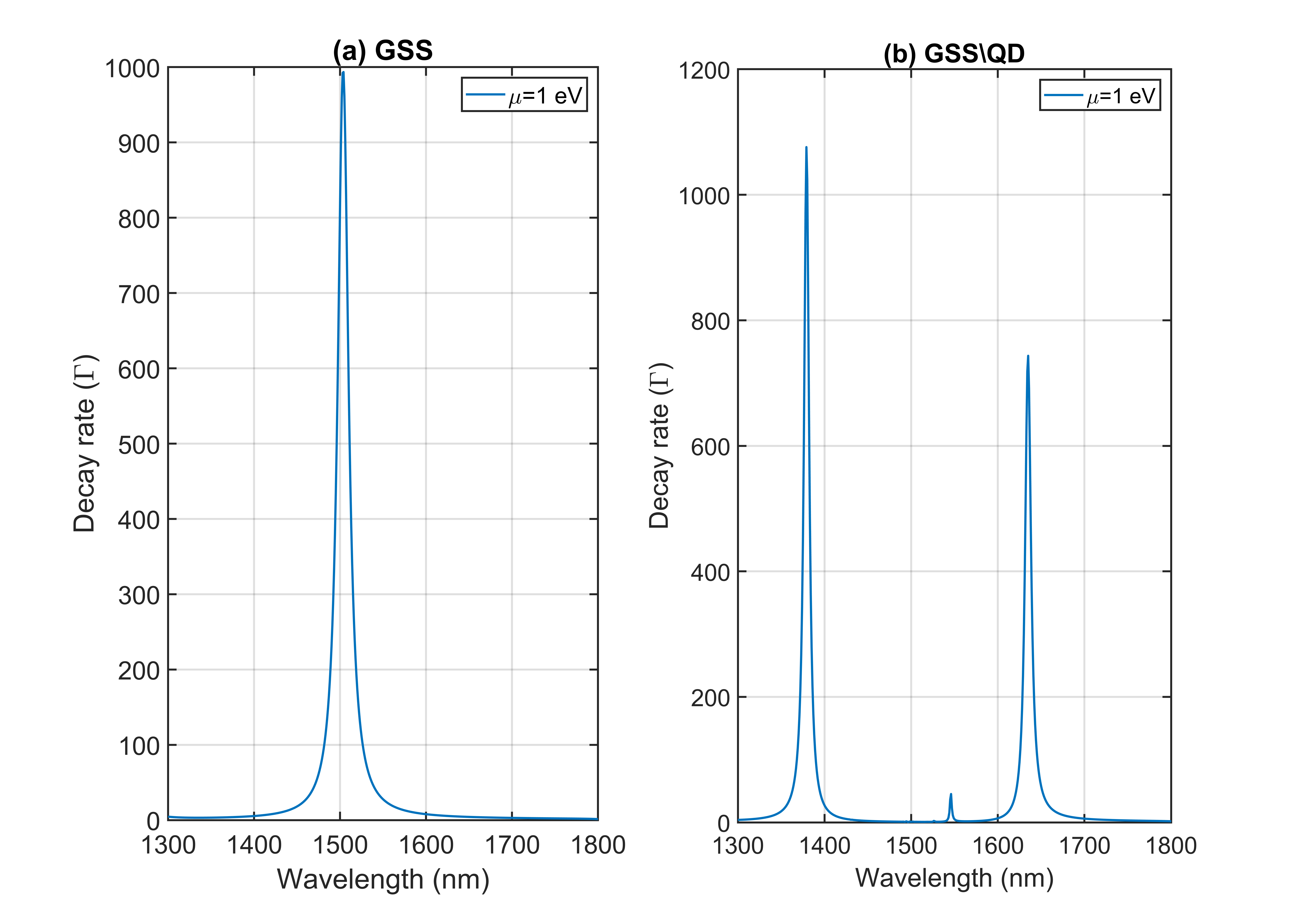}
    \caption{\label{fig:3} Total decay rate of a dipole emitter coupled with (a) GSS and (b) GSS/QD structures as a function of wavelength. The chemical potential of graphene is taken as 1 eV. The transition wavelength ($\lambda_{eg}$) of QD is 1460 nm.}
\end{figure}
where graphene is modeled as a thin layer with thickness d $= 0.5$ nm and surface conductivity $\sigma(\omega)$ is given by \cite{Novoselov2004}. In the random phase approximation, the optical response of graphene is given by its in-plane surface conductivity $\sigma = \sigma_{intra}+\sigma_{inter}$, consisting of intra- and inter-band transitions determined by electron-hole pair excitations. The surface conductivity of graphene depends on the chemical potential ($\mu$), temperature and the scattering energy (E$_s$) values \cite{Wunsch2006}. Here, we vary the value of $\mu$, for the active tuning of graphene \cite{Novoselov2004}. In Fig.\ref{fig:2}, we demonstrate the extinction/absorption spectrum of GSS with and without QD illuminating by plane-wave. The peaks in the absorption spectrum represent localized surface plasmon (LSP) mode excited in GSS. The resonance wavelength ($\lambda$) of LSP mode is evaluated through the following relation,
\begin{equation}
\lambda = 2\pi c \sqrt{\frac{\hbar\epsilon}{\pi a \mu}\frac{1}{12}R}
 \label{eq:2}
\end{equation}
where R is the radius of graphene spherical shell and $\epsilon$ is the dielectric permittivity of the surrounding medium and the space inside the shell. Since the radii of graphene shell is much smaller than the excitation wavelength ($\lambda$) R$\ll \lambda$, therefore we focus on dipole mode only. Moreover, in this regime the extinction and absorption have essentially the same value (therefore we disregard the scattering) \cite{gunay_2020}.\\
\begin{figure}[t!]
      \centering
        \includegraphics[scale=0.28]{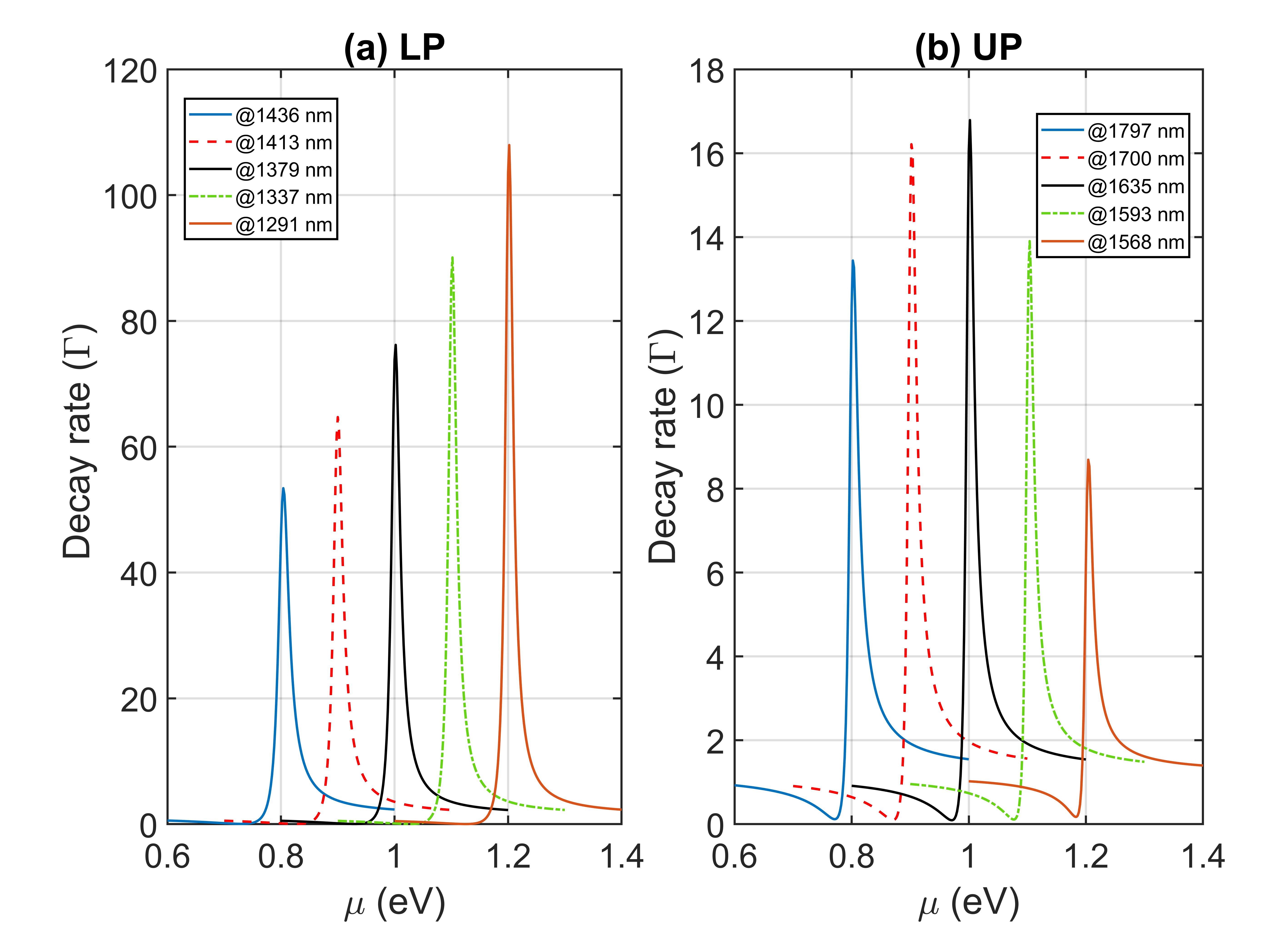}
    \caption{\label{fig:4} Total decay rate of a dipole emitter as a function of $\mu$ for different (a) LP and (b) UP plexcitonic spectral position . The transition wavelength $\lambda_{eg}$ of QD is taken as 1460 nm.}
\end{figure}  
In Fig.\ref{fig:2}(a), the absorption spectrum show the LSP mode resonances of graphene shell in the infrared regime. By applying a small voltage to AFM tip the chemical potential of graphene changes which then blue-shifts LSP resonances to shorter wavelengths. The shift in the spectral position as the chemical potential increases resulted due to increase in the optical gap, as predicted by Eq.(\ref{eq:1}). When a QD of radius 5 nm is coated with GSS, the excitonic states with off-resonant frequency strongly couple to plasmon mode splitting is observed in the absorption spectrum. This splitting is resulted due to the energy exchange between the off-resonant exciton mode and the localized surface plasmon mode of GSS. Moreover, the absorption is enhanced up to 3-fold in the presence of QD. Here, the QD is modeled by Lorentzian dielectric function \cite{Wu:10, Postaci2018}.
\begin{equation}
\epsilon_{eg}(\omega) = \epsilon_{\inf}- f \frac{\omega^2_{eg}}{\omega^2- \omega^2_{eg} + i \gamma_{eg}\omega}
\label{eq:3}
\end{equation}
where $\epsilon_{\inf}$ is the bulk dielectric permittivity at higher frequencies, $f$ is the oscillator strength, $\gamma_{eg}$ and $\omega_{eg}$ are the decay rate and transition frequency of QD, respectively. The QD supports exciton polaritons and in the electrostatic limit this dipole mode is excited with resonant condition $Re\epsilon_{eg}(\omega)= -2\epsilon$, where $\epsilon =1$ is the dielectric permittivity of surrounding medium. Under dipole approximation and resonant condition, changing the radius of the QD does not effect its resonance wavelength, however the dipole resonances changes according to the level-spacing of QD.   
In Fig.\ref{fig:2}(b), we simulate the GSS/QD structure in the same spectral region where the transition wavelength of QD is fixed at $\lambda_{eg}= 1550$ nm which is off-resonant to LSP mode. The graphene LSP resonances split in two plexciton modes as anticipated, the upper plexciton (UP) peaks around 1580 nm to 1800 nm and lower plexciton (LP) ranges from 1290 nm to 1450 nm. For the value of $\mu=0.8$ eV the splitting between UP and LP in the absorption spectrum is 166 meV which increases and blueshifts to 176 meV as chemical potential increases to $\mu=1.2$ eV. The Rabi splitting between graphene plasmons and exciton is stronger even in the off-resonant regime. Tuning these plexciton modes through external voltage provides the leverage to modulate the decay rate dynamics of QE placed close to the GSS/QD structure, provided that if QE emits in the IR wavelengths. Recently, many studies have demonstrated different materials such as hexagonal boron nitride (hBN), two dimensional TMD materials, defect centers and nitrogen vacancies (NV) with emission wavelengths in infrared regime \cite{Zhao2021, Huang2021, Basha2024}. 
\begin{figure}[t!]
      \centering
        \includegraphics[scale=0.32]{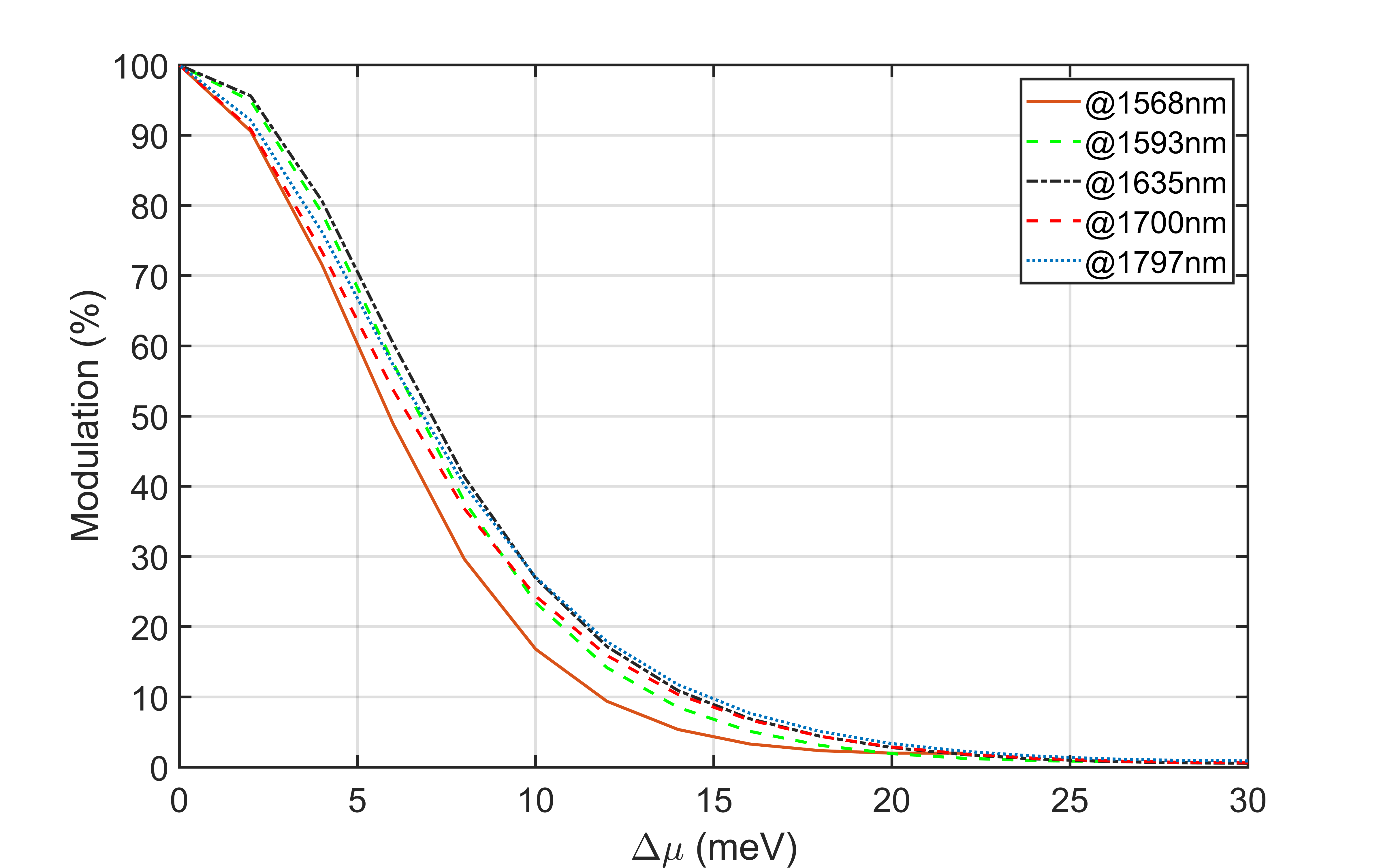}
    \caption{\label{fig:5} Decay rate modulation as a function of chemical shift $(\Delta\mu)$ at different spectral positions of UP.}
\end{figure}
We evaluate the total decay rate of a dipole emitter in TMD layer, placed under the GSS/QD structure, through MNPBEM simulation. The decay rate $\Gamma$ of dipole is proportional to the strength of the coupling between the transition dipole matrix element d and the electromagnetic modes acting on it, including the plasmon. This can be related to the electric field induced by the dipole on itself E$_{ind}$, reflected by the graphene. The exact relation of total decay rate $\Gamma_{tot}$ is given below; 
\begin{equation}
\Gamma_{tot}=\Gamma^o_{rad}+\frac{1}{2}\mathcal{I}m[d. E_{ind}]
\label{eq:4}
\end{equation}
Here, $\Gamma^o_{rad}=4k_o^3|d|^2/3\hbar$ is the free space decay rate \cite{Anger2006}. In Fig.\ref{fig:3}, we evaluate the modulation in the decay rate of dipole emitter with bare GSS structure for $\mu=1$ eV and compared it with the decay rate in the presence of GSS/QD coupled plexcitonic modes. Fig.\ref{fig:3}(a), shows the decay rate of dipole in the presence of hollow graphene shell with peak at 1500 nm. In the case of coupled GSS/QD, the peak splits into two sharp resonances at 1380 nm and 1650 nm, as shown Fig.\ref{fig:3}(b). The splitting of UP and LP modes with bandwidth around 84 meV give the advantage of tuning decay rate at different spectral positions from near- to far-infrared regime. Moreover, due to plexciton coupling the linewidth of dipole at these spectral positions becomes much sharper than coupling with hollow GSS making it more advantageous for sensing applications. Now by applying voltage, the graphene plexciton pronouncedly shift to left and right side with a broad spectral bandwidth due to modulation in $\mu$.\\
We investigate the decay rate modulation at the tuned spectral position of UP and LP modes. Figure.\ref{fig:4}(a) and \ref{fig:4}(b) shows spectral shifts of LP and UP modes as a function of $\mu$, respectively. At a specific wavelength of LP, i.e., 1436 nm the decay rate modifies from 0 to 50 with a small shift in the $\mu$. For instance, a 40 meV shift in chemical potential yields a two-orders of magnitude of increase in decay rate at spectral wavelength of 1379 nm. Similar trend is observed for the peaks of UP in Fig.\ref{fig:4}(b). To demonstrate this we plot the modulation in the decay rate at specific wavelengths of UP as a function of chemical shift, as shown in Fig.\ref{fig:5}. The graph indicates that maximum modulation in the decay rate occur with a very small shift in the chemical potential. This continuous shift in the decay rate from maximum to minimum paves the way for controlled spontaneous photon emission, ultrafast switching and compact platforms for tunable single photon sources.\\

\section{Conclusion}
In this study, we demonstrated active tuning of the decay rate of a dipole through graphene spherical shell and QD coupled system. In a table top setup, we electrically modulated the chemical potential of graphene shell which results not only in the broadband spectral modulation of plexcitonic modes but also significantly modifies the decay rate of dipole emitter placed in the close proximity of GSS-QD structure. The plasmon-exciton strong coupling demonstrate sharp linewidths of plexcitons even in the off-resonant regime, which indicates high sensitivity of photonic devices at different wavelengths. Furthermore, controlling decay rate of a dipole through graphene plexciton provides the leverage to control the emission properties from mid-infrared to far-infrared regime. This controlled modulation of dipole's decay rate opens substantial opportunities for practical applications, including quantum technologies, quantum sensing, THz telecommunications and quantum networks. Our findings offer significant insights into the control of spectral and temporal dynamics of dipole emitter by the application of a small voltage, thereby contributing robustly to the field of on-demand graphene-based integrated photonic systems and quantum information processing.

\begin{acknowledgments}
R.S., T.T.A., and H.A. acknowledge support from TUBITAK No. 123F156.
\end{acknowledgments}
\normalsize $^{\dagger}$These authors equally contributed to this work. 
\nocite{*}
\bibliography{GDR}

\end{document}